# The General sampling theorem, Compressed sensing and a method of image sampling and reconstruction with sampling rates close to the theoretical limit

L. Yaroslavsky
School of Electrical Engineering, Tel Aviv University, Tel Aviv, Israel

E-mail: yaro@eng.tau.ac.il

## Abstract

The article addresses the problem of image sampling with minimal possible sampling rates and reviews the recent advances in sampling theory and methods: modern formulations of the sampling theorems, potentials and limitations of Compressed sensing methods and a practical method of image sampling and reconstruction with sampling rates close to the theoretical minimum.

## 1. Introduction.

Sampling is the very first step in digital imaging. The fundamental part of its theoretical base is sampling theory. The origins of the sampling theory date back to 1920-1940th years to classical publications by H. Nyquist, V. A. Kotelnikov and C. Shannon ([ 1], [ 2], [ 3]). The classical theory is based on the concept of band-limited signals, i.e. signals whose Fourier spectrum is non-zero only within a finite frequency interval in the signal Fourier domain. In the last two decades, the needs of the development of digital imaging engineering inspired new advances in the sampling theory associated with notions of signal spectrum sparsity, Compressed sensing, and methods of signal sampling with sampling rates close to the theoretical limits ([ 4], [ 5], [ 6], [ 7], [ 8], [ 9],[ 10], [11], [ 12], [ 13], [14], [ 15]).

The article represents a brief review of recent publications on these advances. Section 2 presents classical and modern formulations of the sampling theorem. In Section 3 the General sampling theorem is formulated and minimal sampling rate is evaluated using the concept of signal sub-band decomposition. Section 4 presents an alternative derivation of the General sampling theorem based on a discrete signal model and The Discrete sampling theorem. In Section 5 the ubiquitous compressibility of digital images acquired by conventional imaging devices is discussed. Section 6 briefly reviews potentials and limitations of the Compressed sensing methods advanced as a solution to the problem of minimization of the signal sampling rates. In Section 7 a new method of image sampling and reconstruction is outlined that allows reaching sampling rates close to the theoretical minimum. In the concluding Section 8 some practical issues of implementation of this method and its possible applications for solving other imaging problems are discussed.

## 2. The classical sampling theorem

The classical Kotelnikov-Shannon 1D sampling theorem ([ 2], [ 3]) states that band-limited signals can be precisely reconstructed from their samples taken with sampling interval $\Delta = 1/F$ one from another, where $[-F/2, F/2]$ is the interval in the frequency domain of the signal Fourier Transform, that contains the entire signal spectrum.

1D band-limited signals with Fourier spectrum concentrated within a bounded interval $\left[-F/2, F/2\right]$ around zero frequency are called baseband signals. 1D band-limited signals with spectrum concentrated within intervals $\left[f_0 - F/2, f_0 + F/2\right]$ and $\left[-f_0 - F/2, -f_0 + F/2\right]$ around a non-zero frequency $f_0$, called the carrier frequency, are called passband signals. According to the properties of the Fourier transform, a passband signal $a_{PB}(x)$ can be regarded as a result of modulation of a baseband signal $a_{BB}(x)$ by a sinusoidal signal of frequency $f_0$:

$$a_{PB}(x) = a_{BB}(x)\sin(2\pi f_0 x) \qquad (1)$$

The classical sampling theorem can be straightforwardly applied to passband signals if, before sampling, passband signals are converted into the corresponding baseband signals by multiplying, or demodulating, them by a sinusoidal signal of the carrier frequency $f_0$ and subsequent ideal low-pass filtering of the demodulation result within the baseband $\left[-F/2, F/2\right]$.

For 2D signals, such as images, the classical sampling theorem states that signals with Fourier spectrum band-limited by a rectangular interval, say $\left[-F_x/2, F_y/2\right]\left[-F_x/2, F_y/2\right]$, in the frequency domain $\left(f_x, f_y\right)$ of the signal Fourier Transform can be precisely reconstructed from their samples taken with sampling intervals $\left[\Delta_x = 1/F_x, \Delta_y = 1/F_y\right]$ one from another at nodes of the rectangular sampling lattice in signal Cartesian coordinates $(x, y)$.

In reality no band-limited signals exist and only approximate reconstruction of signals from their sampled representation is possible. In view of this, the classical sampling theorem should be reformulated in terms of signal band-limited approximation with a given mean square error (MSE):

*Signal $a(x, y)$ sampled with sampling intervals $\left(\Delta_x, \Delta_y\right)$ over the rectangular sampling lattice can be reconstructed from its samples with no distortions caused by spectra aliasing due to sampling and with MSE equal to the signal energy (integral of signal spectrum square module) outside the rectangular frequency interval $\Omega_{rect} = \left|-1/2\Delta_x, 1/2\Delta_x; -1/2\Delta y, 1/2\Delta_y\right|$ called the signal sampling base band, iff signal sampling and reconstruction are carried out using sampling and reconstruction devices with frequency responses $Fr^{(samp)}\left(f_x, f_y\right)$ and*

$Fr^{(rec)}\left(f_x, f_y\right)$, *correspondingly, of the ideal low-pass filters :*

$$Fr^{(samp)}\left(f_x, f_y\right) = \begin{cases} \dfrac{1}{\Delta_x \Delta_y}, \left(f_x, f_y\right) \in \Omega_{rect} \\ 0, \qquad otherwise \end{cases};$$

$$Fr^{(rec)}\left(f_x, f_y\right) = \begin{cases} 1, \left(f_x, f_y\right) \in \Omega_{rect} \\ 0, \qquad otherwise \end{cases} \qquad (2)$$

## 3. The General sampling theorem. A formulation based on the concept of signal sub-band decomposition

The classical 2D sampling theorem implies that the minimal signal sampling rate, i.e. the minimal number of samples per unit of the signal area sufficient for signal reconstruction with a given MSE equals the area $F_x \times F_y = 1/\Delta_x \times 1/\Delta_y$ of the 2D rectangular band-limiting frequency interval $\Omega_{rect}$ that contains signal frequency components, which all together reproduce the signal with this MSE.

Generally, the signal frequency interval $\Omega_{MSE}$ of the minimal area that contains the signal largest frequency components, which reproduce the signal with a given MSE, may have an arbitrary shape. We will call this frequency interval "signal spectrum MSE-defined (MSED-) zone". The General sampling theorem extends the above statement on the signal minimal sampling rate to the general case of arbitrary signals and states:

*The minimal number of samples per unit of signal area sufficient for signal reconstruction from them with a given MSE equals the area of the signal Fourier spectrum MSED-zone.*

The theorem can be proved using the concept of signal sub-band decomposition ([16], [17],). In the sub-band decomposition, image $a(x, y)$ is decomposed into a sum

$$a(x, y) \cong \sum_k a^{(k)}(x, y) \qquad (3)$$

of a certain number $K$ components $\left\{a^{(k)}(x, y)\right\}$ with spectra of rectangular shapes that all together approximate the image spectrum as it is illustrated in Figure 1. Each of the components can be, according to the classical sampling theorem for band-limited and passband signals, reconstructed from its corresponding samples taken with sampling rate $\left\{SR_k\right\}$ equal to its corresponding area $\left\{S^{(k)}_{f_x, f_y}\right\}$. Hence the overall sampling rate $SR_K$ sufficient for precise reconstruction of all $K$ signal sub-bands amounts to

$$SR_K = \sum_{k=1}^{K} SR_k = \sum_{k=1}^{K} S_{f_x, f_y}^{(k)} \qquad (4)$$

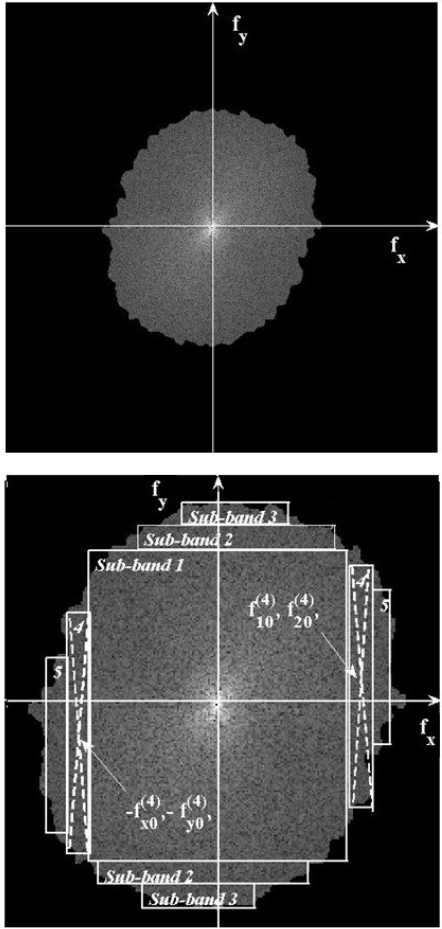

Figure 1. An example of an image spectrum MSED-zone (upper) and its sub-band decomposition (bottom). Carrier frequencies ($f_{x0}^{(4)}, f_{y0}^{(4)}$) for the passband components are indicated only for the fourth component in order to not overburden the image.

In the limit, when $K \to \infty$, the sub-band components cover the entire area $S_{f_x, f_y}$ of the image spectrum and therefore the overall sampling rate amounts to

$$\lim_{K \to \infty} SR_K = S_{x,y} \lim_{K \to \infty} \sum_{k=1}^{K} S_{f_x, f_y}^{(k)} = S_{f_x, f_y} \qquad (5)$$

i.e. to the area $S_{f_x, f_y}$ occupied by the signal Fourier spectrum MSED-zone. Note that, in distinction from the conventional sampling, the signal sampled representation obtained by sampling signal sub-band components consists of samples of signal sub-band components rather than of samples of the signal itself.

## 4. The General sampling theorem. A formulation based on a discrete signal model

Sampling is a special case of signal discretization methods ([16]). In general, discrete representation of signals is obtained as a set of coefficients of signal expansion over a set of discretization basis functions and signal reconstruction from its discrete representation is performed using reconstruction basis functions reciprocal to the discretization ones. Consider this general case using a discrete model.

Let $\mathbf{A}_N$ be a vector of $N$ samples $\{a_k\}_{k=0,1,\ldots,N-1}$ of a discrete signal, $\mathbf{\Phi}_N$ be an $N \times N$ orthonormal transform matrix

$$\mathbf{\Phi}_N = \{\varphi_r(k)\}, k = 0,1,\ldots,N-1, \ r = 0,1,\ldots,N-1 \qquad (6)$$

composed of basis functions $\{\varphi_r(k)\}$ and $\mathbf{\Gamma}_N$ be a vector of signal transform coefficients $\{\gamma_r\}$ such that:

$$\mathbf{A}_N = \{a_k\} = \mathbf{\Phi}_N \mathbf{\Gamma}_N = \left\{ \sum_{r=0}^{N-1} \gamma_r \varphi_r(k) \right\}. \qquad (7)$$

Select a subset $\widetilde{\mathbf{R}}$ of $K$ transform coefficients indices $\{\widetilde{r} \in \widetilde{\mathbf{R}}\}$ and define a "$K$ of $N$"-bounded spectrum (BS-) approximation $A_N^{(BS)}$ to the signal $\mathbf{A}_N$ as:

$$\mathbf{A}_N^{BS} = \left\{ a_k^{BS} = \sum_{\widetilde{r} \in \mathbf{R}} \widetilde{\gamma}_{\widetilde{r}} \varphi_{\widetilde{r}}(k) \right\} \qquad (8)$$

Assume that available are only $K < N$ samples $\{a_{\widetilde{k}}^{BS}\}_{\{\widetilde{k}\} \in \widetilde{\mathbf{K}}}$ of this signal, where $\widetilde{\mathbf{K}}$ is a $K$-size subset of indices $\{\widetilde{k}\}$ from the set $\{0,1,\ldots,N-1\}$. These available $K$ signal samples define a system of $K$ equations:

$$\left\{ a_{\widetilde{k}}^{BS} = \mathbf{\Phi}_{KofN} \cdot \widetilde{\mathbf{\Gamma}}_K = \sum_{r=0}^{N-1} \gamma_{\widetilde{r}} \varphi_{\widetilde{r}}(\widetilde{k}) \right\}, \ \{\widetilde{k}\} \in \widetilde{\mathbf{K}}, \qquad (9)$$

where $K \times K$ sub-transform matrix $\mathbf{\Phi}_{KofN}$ is composed of samples $\varphi_{\widetilde{r}}(\widetilde{k})$ of the basis functions with indices $\{\widetilde{r} \in \widetilde{\mathbf{R}}\}$ for signal sample indices $\widetilde{k} \in \widetilde{\mathbf{K}}$, and $\widetilde{\mathbf{\Gamma}}_K$ is a vector composed of the corresponding subset $\{\gamma_r\}$ of the signal transform coefficients. One can find these coefficients by inverting matrix $\mathbf{\Phi}_{KofN}$

$$\widetilde{\mathbf{\Gamma}}_K = \{\gamma_{\widetilde{r}}\} = \mathbf{\Phi}_{KofN}^{-1} \cdot \widetilde{\mathbf{A}}_K \qquad (10)$$

provided that matrix $\mathbf{\Phi}_{KofN}^{-1}$ inverse to the matrix $\mathbf{\Phi}_{KofN}$

exists. The latter is conditioned by positions $\left\{ \widetilde{k} \in \widetilde{\mathbf{K}} \right\}$ of available signal samples and by the selection of the subset $\left\{ \widetilde{R} \right\}$ of transform basis functions.

Found in this way transform coefficients together with the rest of the coefficients set to zero can be used for obtaining a bounded spectrum (BS-) approximation $\mathbf{A}_N^{BS}$ to the complete signal $\mathbf{A}_N$ with the mean square error:

$$MSE = \left\| A_N - \hat{A}_N \right\|^2 = \sum_{k=0}^{N-1} \left| a_k - a_k^{BS} \right|^2 = \sum_{r \notin \widetilde{R}} \left| \gamma_{\widetilde{r}} \right|^2 \quad (11)$$

This error can be minimized by an appropriate selection of $K$ basis functions of the sub-transform $\mathbf{\Phi}_{KofN}$. In order to do so, one should know the energy compaction ordering of the basis functions of the transform $\mathbf{\Phi}_N$, i.e. the order of basis functions, in which the energy (squared module) of signal representation coefficients decays with their indices. If, in addition, one knows a transform that is capable of the best energy compaction into the smallest number of transform coefficients, one can, by choosing this transform, secure the best minimum mean square error bounded spectrum approximation of the signal $\left\{ a_k \right\}$ for the given subset $\left\{ a_k^{BS} \right\}$ of its samples. The subset $\mathbf{\Omega}_{MSE}$ of indices $\left\{ \widetilde{r} \right\}$ of the largest transform coefficients, which reconstruct the signal with a given *MSE* is the above-mentioned signal spectrum MSE Defined (MSED-) zone.

With the above reasoning, on can formulate in two statements the following Discrete Sampling Theorem ([14], [ 15], [ 16], [17]):

*Statement 1. For any discrete signal of $N$ samples defined by its $K \leq N$ samples, its bounded spectrum, in terms of a certain transform $\mathbf{\Phi}_N$, approximation can be obtained with mean square error defined by Eq. 11 provided positions of the samples secure the existence of the matrix $\mathbf{\Phi}_{KofN}^{-1}$ inverse to the sub-transform matrix $\mathbf{\Phi}_{KofN}$ that corresponds to the spectrum bounding. The approximation error can be minimized by using a transform with the best energy compaction capability.*

*Statement 2. Any signal of $N$ samples that is known to have only $K \leq N$ non-zero transform coefficients for a certain transform $\mathbf{\Phi}_N$ ( $\mathbf{\Phi}_N$ -transform "bounded spectrum" signal) can be precisely reconstructed from exactly $K$ its samples provided positions of the samples secure the existence of the matrix $\mathbf{\Phi}_{KofN}^{-1}$ inverse to the sub-transform matrix $\mathbf{\Phi}_{KofN}$ that corresponds to the spectrum bounding.*

Potential candidates of signal transforms with good energy compaction capability that can be used for obtaining signal bounded spectrum approximations are Discrete Fourier Transform (DFT), Discrete Cosine Transform (DCT), Walsh Transform and Wavelet Transforms. DFT and DCT admit arbitrary positions of the available signal samples, whereas Walsh Transform and Wavelet Transforms impose on positioning signal samples certain limitation ([18]).

In the limit, when the number of samples $N$ in our discrete model tends to the infinity, the model transmutes to a continuous one, discrete Fourier transform transmutes to the integral Fourier transform and the Discrete Sampling Theorem for the DFT transmutes to the above-formulated General Sampling Theorem.

## 5. The ubiquitous compressibility of images sampled over the standard rectangular sampling lattices

Sampling images over the uniform rectangular sampling lattices assumed by the classical 2D sampling theorem is accepted as the standard in imaging engineering for designing image scanners, digital cameras and image display devices. It also is assumed by default by image processing software. This by the necessity implies that in order to avoid image distortions additional to those that define the image spectrum MSED-zone, image MSED-zones must be inscribed into the sampling base-band rectangle defined by the image sampling rate. Therefore sampling images over the rectangular sampling lattices always requires sampling rates that exceed the minimal one defined by the area of the image spectrum MSED-zone, i.e. images sampled in the standard way are always oversampled. This causes the need in image compression for their storage and transmission. The degree of the over-sampling, or the oversampling redundancy, is inverse to the ratio of the MSED-zone area to the area of the image sampling rectangular baseband. This ratio is called the image spectrum sparsity.

Image spectrum sparsity is illustrated in Figure 2, which presents an example of a test image and its Fourier spectrum centred at zero spatial frequencies. Highlighted in the spectrum is the image spectrum MSED-zone that contains image spectral components, which reconstruct the image with MSE set, in this particular example, to be equal to that of the image JPEG compression by Matlab means. Image sampling baseband in this spectrum is the entire area of the image spectrum. Spectrum sparsity of this image is 0.31. Experimental experience evidences that spectra sparsities of sampled natural images lie usually in the range 0.1-0.4.

**Test image of 512x512 pixels**

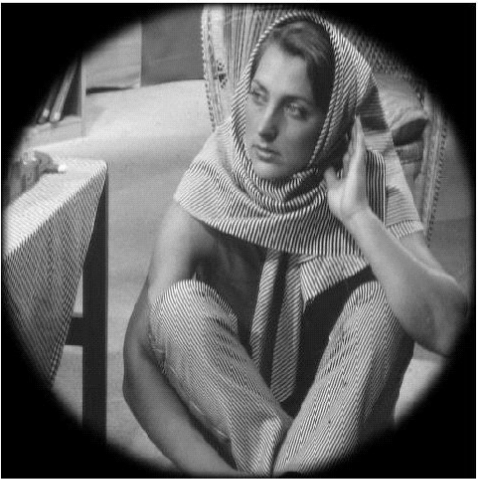

**Image spectr&Spectr mask. Enrg thresh (Jpg)=0.997**
**ErrStd=3.62; SpectrSparsity=0.31**

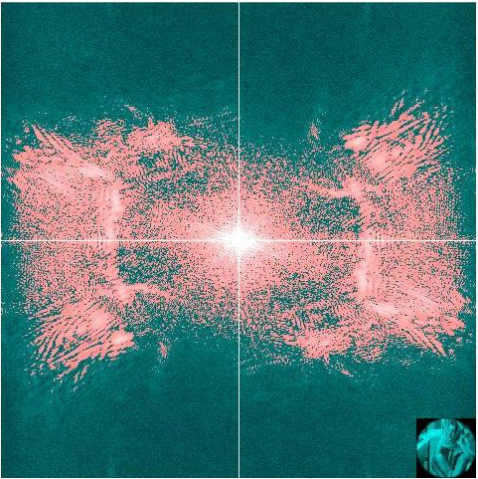

Figure 2. An example of a test image (upper) and its Fourier spectrum (bottom) centred at its DC component. Highlighted (by red in e-version) in the spectrum are the largest spectral components sufficient for image reconstruction with the same MSE as that of its JPEG compression (the image spectrum MSED-zone).

## 6. Compressed sensing: potentials and limitations

The ubiquitous compressibility of images acquiered by the conventional methods raises a very natural question: "is it possible to just directly measure the minimal amount of data and avoide the need in image compression? " This question was first posed by the inventors of the Compressed sensing approach (known also under the name "Compressed sampling") as a solution to this problem ([ 4]**,** [ 5]**,** [ 6]**)**.

The Compressed sensing approach considers signal sampling and reconstruction as an underdetermined inverse problem of recovering a signal of $N$ samples from a fewer number $K < N$ of measurements ([ 4],[ 5], [ 6]). Because the signal recovering problem is underdetermined, it has an indefinite number of solutions unless there is an a priori restriction on signals to be recovered that allows choosing from all possible solutions the only one that satisfies this restriction. In the Compressed sensing approach, this restriction is formulated in terms of **L0** norm in spectral domain of a certain image transform, i.e. in terms of the amount of signal non-zero transform coefficients.

According to the theory of the Compressed sensing, if an image of $N$ samples is known to have only $K < N$ non-zero transform coefficients in the domain of a certain "sparsifying" transform, it can be precisely reconstructed from a certain number $M > K$ measurements by means of minimization of the **L0** norm in the image transform domain ([ 4], [ 5[ 5]). This, in particular, means that signal sampling rate, in distinction from the conventional sampling according to the classical sampling theorem, does not depend on the signal highest frequency and, in particular, that the signal sampling rate can be lower than twice the signal highest frequency, i.e. a kind of a "sub-Nyquist" sampling with aliasing is admissible.

False spectra aliasing components within the sampling base band caused by the "sub-Nyquist" signal sampling can not, in principle, be filtered out by signal linear filtering. The minimization of **L0** norm of image spectra suggested by the Compressed sensing approach implements a kind of a nonlinear filtering for separating true signal spectrum components from the false aliasing ones.

The capability of the Compressed sensing approach of reconstructing signals sampled with aliasing can be demystified using the following simple model of sampling and reconstruction of signals composed from a known number of sinusoidal components ([ 13]).

Let $N$ be a number of signal samples, $K < N$ be a number of its sinusoidal components and let the signal be sub-sampled at $M > K$ arbitrarily chosen points. It is required to precisely reconstruct all $N$ signal samples from these $M$ available samples. This would be achieved if one could determine amplitudes and frequencies of the signal sinusoidal components. Figure 3 and Figure 4 illustrate, utilizing results of a computer simulation, how and when this can be done.

A test signal presented in Figure 3, 1st plot from the top, has five sinusoidal components ( $K = 5, N = 512$ ) seen as five Kronecker deltas in the signal Discrete Cosine Transform spectrum (2nd plot from the top in Figure 3). When this signal is sub-sampled, in its spectrum (3rd plot in Figure 3) a lot of false aliasing spectral components appear. In the shown example the signal is sub-sampled at $M = 76$ random positions. In this particular example the signal's five true spectral components exceed the aliasing ones and are easily detectable by finding positions of the given number $K = 5$ the largest spectral components.

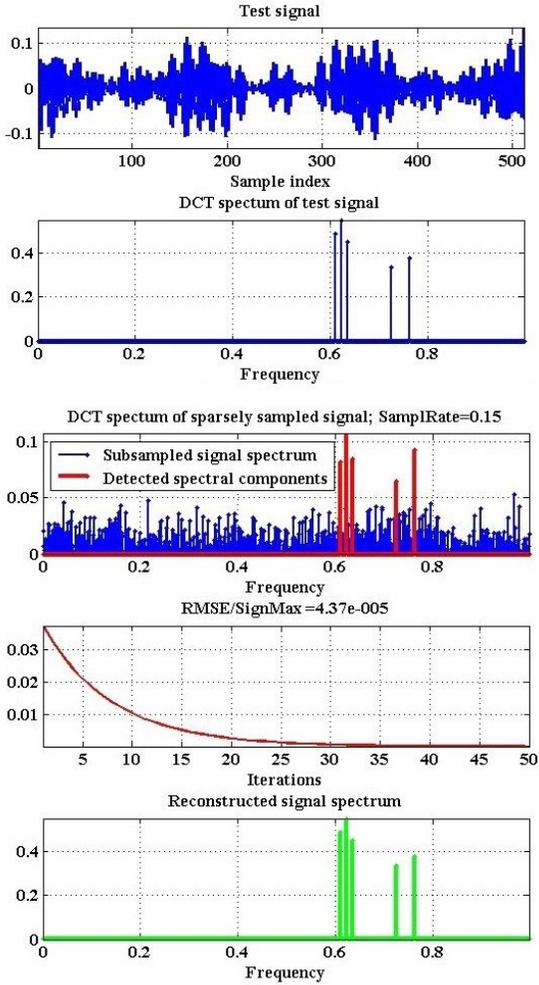

Figure 3. Reconstruction of a test signal sampled with aliasing. From top to bottom: a test signal composed of five sinusoidal components; test signal DCT spectrum; DCT spectrum of the sub-sampled signal; plot of the root mean square reconstruction error normalized to the signal maximum (RMSE/SignMax) vs. the number of the reconstruction iterations; DCT spectrum of the reconstructed signal. Frequency in plots of spectra is given in fractions of the sampling base band.

Once this is done, reconstruction of all signal $N$ samples can be achieved using an iterative Gerchberg-Papoulis type algorithm, in which, at each iteration step, (i) DCT of the current estimate of the reconstructed signal is computed; (ii) positions of the given number $K$ the largest spectral components are detected; (iii) all spectral components except the detected ones are set to zero; (iv) inverse DCT of the modified in this way signal spectrum is computed; (v) available signal samples are restored in the obtained estimate of the reconstructed signal to get a next estimate of the reconstructed signal and the loop is repeated. The plot of the reconstruction root mean square error (RMSE) normalized to signal maximum vs. the number of reconstruction iterations (4-th plot in Figure 3) and the plot of DCT spectrum of the reconstructed by the iterative algorithm signal (5-th plot) illustrate this process and demonstrate that practically precise reconstruction of the signal is achieved.

Obviously, when the signal sub-sampling rate is not sufficiently high and spectrum aliasing is severe, reliable detection of the signal true spectral components in the spectrum of the sampled signal and, hence, signal reconstruction become impossible. This case is illustrated in Figure 4 by the results of the same experiment and with the same parameters as in Figure 3, but for another realization of positions of signal samples.

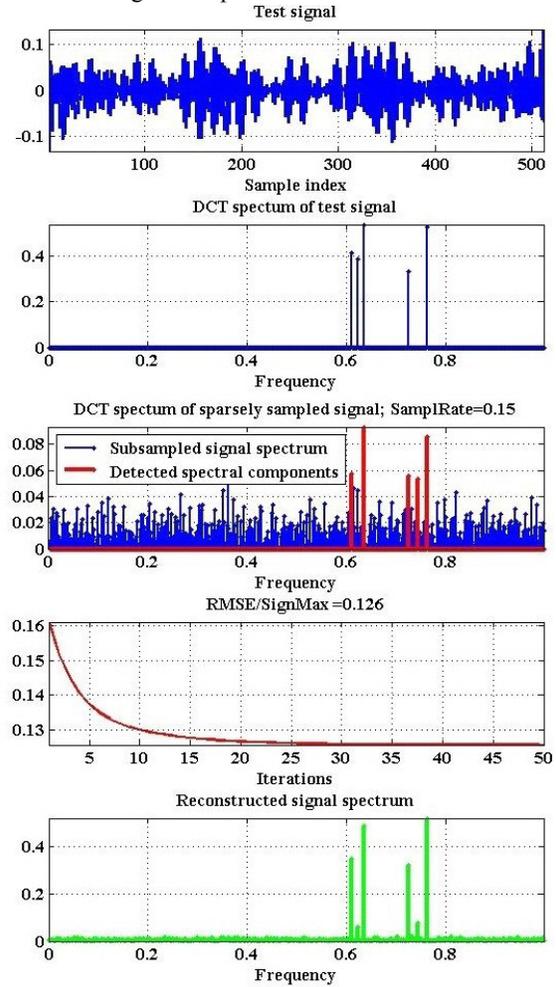

Figure 4. Failure of reconstruction of a test signal sampled with aliasing. From top to bottom: a test signal composed of five sinusoidal components; test signal DCT spectrum; DCT spectrum of the sub-sampled signal; plot of the root mean square reconstruction error normalized to the signal maximum (RMSE/SignMax) vs. the number of the reconstruction iterations; DCT spectrum of the reconstructed signal. Frequency in plots of spectra is given in fractions of the sampling base band.

In this case two of the signal spectrum components are not detected and false peaks are detected instead. As a result, signal reconstruction by the iterative algorithm failed.

In the presented examples, signal spectrum sparsity, i.e. the ratio of the number of the signal non-zero spectral coefficients to the total number of spectral coefficients, is $K/N = 5/512 \approx 10^{-2}$. According to the Discrete sampling theorem the minimal number of signal samples needed for its precise reconstruction is $K = 5$, whereas actually $M = 76$ samples were required for signal reconstruction in the example of the successful signal reconstruction shown in Figure 3. Therefore, the sampling redundancy, i.e. the ratio of the actual number of signal samples required in this case for signal reconstruction to the number of the signal non-zero spectral coefficients, is $R = M/K = 76/5 \cong 15.2$.

Two shown examples demonstrate that when sampling is performed in random positions, signal precise reconstruction is possible with a certain probability depending on the sampling redundancy. Figure 5 presents results of an experimental evaluation of the sampling redundancy required by the above-described algorithm. The results were obtained by Monte-Carlo simulation of sampling and reconstruction using a model of a single sinusoidal signal. The experiments were conducted for sinusoidal signals ($K = 1$) of five different frequencies (0.9, 0.7, 0.5, 0.3 and 0.1 of the signal base band) and of five different signal lengths $N$ (128, 256, 512, 1024, 2048), i.e. of five different signal sparsities $K/N$.

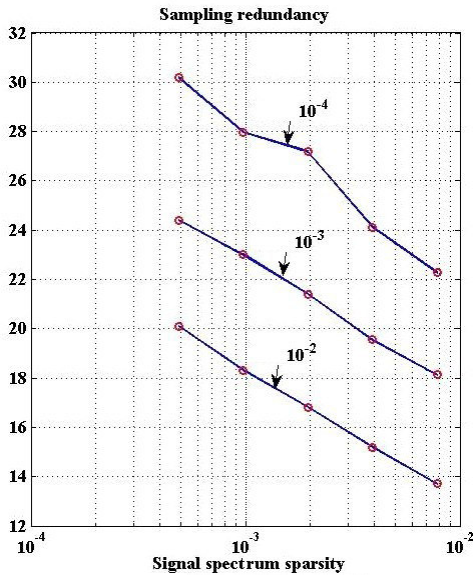

Figure 5. Estimates of the sampling redundancy required for reconstruction of sinusoidal signals from their randomly placed samples vs. the signal sparsity for three probabilities of the reconstruction failure ($10^{-2}$, $10^{-3}$, and $10^{-4}$).

They show that sampling redundancies in the range of 14-32 times are required for signal reconstruction with the probability of signal frequency identification error less than $10^{-4}$, $10^{-3}$ and $10^{-2}$, correspondingly. These results were obtained over $5 \times 10^4$ realizations of the random sampling for each individual experiment with a given sampling rate, signal frequency and signal length.

From this illustrative model one can also conclude that the presence of noise in the sampled data will hamper reliable detection of the signal spectral components and will require an additional sampling redundancy for the signal reconstruction.

Potentials of the Compressed sensing approach to image sampling and reconstruction are widely advertised in the literature. Much less is known about its limitations. Particularly important is the question, how close is the amount of measurements required by the Compressed sensing methods of image sampling and reconstruction to the theoretical minimum defined by the sampling theory.

According to the theory of the Compressed sensing, the precise reconstruction of a signal of $N$ samples that has $K < N$ non-zero transform coefficients is possible, when the number of measurements $M$ sufficient for signal reconstruction satisfies the following inequality ([ 8],[ 9])

$$M/K > -2\log\left[(M/K)(K/N)\right] \qquad (12)$$

By virtue of the Discrete sampling theorem, the signal sparsity $SS = K/N$ is the theoretical lower bound of the sampling rate required for signal reconstruction. Therefore the ratio $R = M/K$ represents the sampling redundancy with respect to the theoretical minimum. Inequality (12) can be rewritten as a relationship between the signal sparsity $SS = K/N$ and the sampling redundancy $R = M/K$ as

$$R > -2\log(R \times SS) \qquad (13)$$

Numerical evaluation of this relationship between the sampling redundancy $R$ and the signal sparsity $SS$ gives that in the above-mentioned range from 0.1 to 0.4 of spectra sparsities of natural images the sampling redundancy of the Compressed sensing methods should theoretically be larger than 2 to 3. Experimental data collected over publications show that in practice it should be larger than 2.5 to 5 ([11]). This means that the sampling redundancy required by the Compressed sensing methods for natural images is of the same order as the sampling redundancy of their regular sampling (2.5 - 10), i.e. in reality Compressed sensing methods do not solve the problem of the compressibility of images acquired by the conventional means.

The substantial sampling redundancy required by the Compressed sensing methods is not their only drawback. Their applicability is also impeded by the vulnerability to noise in the sensed data and by the impossibility to predict and secure the resolving power of the reconstructed images. Resolving power of images is determined by the size and shape of the MSED-zones of their spectra. Spectra MSED-zones of images reconstructed by the methods of Compressed sensing are formed in the process of image reconstruction rather than are specified in advance from the requirements to the image resolving power.

To summarize the said, Compressed sensing methods are to a certain degree capable of reconstructing sparse approximations of images sampled with aliasing. No a priori knowledge regarding MSED-zones of the image spectra is required for this. This is an attractive feature of the Compressed sensing methods. It however has its price: because of this Compressed sensing methods require a significant redundancy in the number of measurements sufficient for image reconstruction.

In many practical tasks of digital image acquisition, the assumption of the complete uncertainty regarding the image spectra MSED-zones has no justification. In fact, if one is ready, as it is assumed by the Compressed sensing approach, to accept a sparse spectrum approximation to an image and has chosen an image sparsifying transform, one tacitly implies certain knowledge of the energy compaction capability of the chosen transform. Making use of this in any case available a priori knowledge allows implementation of image sampling with sampling rates close to the theoretical minimum.

## 7. A method of image sampling and reconstruction with sampling rates close to the theoretical minimum

The Discrete Sampling Theorem implies that for image sampling and reconstruction one should ([ 12], [ 13], [14], [ 15]):

- Choose an image sparsifying transform that features the best, for the given image, energy compaction capability.
- For a given number $N$ of image samples, specify a desired MSED-zone of the image spectrum, i.e. a set of $M < N$ indices of transform coefficients to be used for image reconstruction.
- Take $M$ image samples.
- Use the obtained $M$ image samples for determining $M$ transform coefficients that belong to the chosen MSED-zone.
- Set the rest $N - M$ transform coefficients to zero and use the obtained spectrum for reconstruction of the required $N$ image samples by its inverse transform.

Consider possible ways for implementation of this protocol.

- <u>Choosing an image sparsifying transform</u>. The key role in the choice of the image sparsifying transform plays the energy compaction capability of the transform. An additional feature, which is usually required, is the availability of a fast transform algorithm. From this viewpoint, Discrete Cosine (DCT), Discrete Fourier (DFT) and Wavelet transforms appear most suitable.

- <u>Defining the image spectrum MSED-zone.</u> Definition of the image spectrum MSED-zone, i.e. indices of transform coefficients to be used for image reconstruction, is based on the known energy compaction capability of the transform. In most cases, DCT can be recommended as the image sparsifying transform.

DCT is known to efficiently compact image largest transform coefficients into quite tight groups in the area of low spatial frequencies. An additional advantage of using DCT as the image sparsifying transform is that it is a version of the discrete representation of the integral Fourier transform and, therefore, it perfectly concords with treatment of imaging systems in terms of their frequency transfer functions [ 16]).

Of course, given an image to be sampled, one can't not precisely specify MSED-zone of its spectrum. However, a considerable practical experience, and, in particular, results of developing zonal quantization tables for JPEG image compression, show that MSED-zones of spectra of natural images are sufficiently well concentrated and can be with a reasonable accuracy circumscribed by one of some standard shapes specified by few geometrical parameters such as total area, angular orientation, aspect ratio, etc. Examples of such standard shapes well suited for approximating MSED-zones of DCT spectra of natural images are presented in Figure 6. One can associate each particular shape with a certain class of images such as micrographs, aerial photographs, space photos, in-door and out-door scenes, etc.

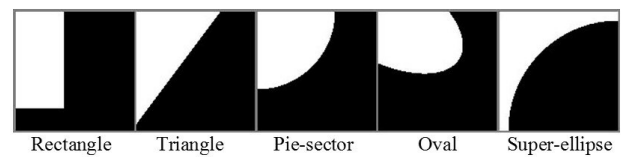

| Rectangle | Triangle | Pie-sector | Oval | Super-ellipse |

Figure 6. Possible standard shapes for approximating the MSED-zones of image DCT spectra (spectra DC components are in the upper left corners of the shapes).

Standard shapes for approximating signal spectra MSED-zones have two important properties: (i) they do not require fine tuning of their geometrical parameters to fit the spectra MCSED-zones they are chosen to approximate and (ii) their areas by the necessity always exceed the areas of their corresponding image spectra MSED-zones. These properties are illustrated in Figure 7. The spectrum MSED-zone of a test image (Figure 7, a) is shown in figures b)-e) by white dots. It is obtained for the image reconstruction root mean square error (RMSE) 3.85 gray levels of 256 gray levels for the image dynamic range, the same as the reconstruction RMSE of image JPEG compression by Matlab means.

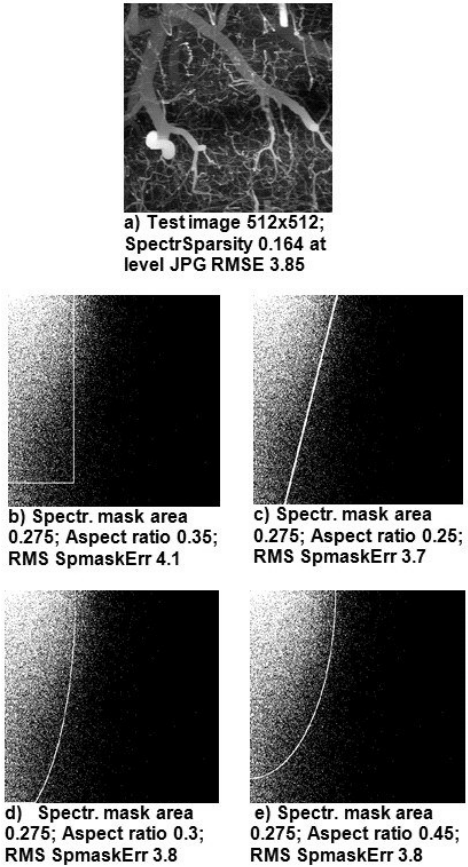

a) Test image 512x512; SpectrSparsity 0.164 at level JPG RMSE 3.85

b) Spectr. mask area 0.275; Aspect ratio 0.35; RMS SpmaskErr 4.1

c) Spectr. mask area 0.275; Aspect ratio 0.25; RMS SpmaskErr 3.7

d) Spectr. mask area 0.275; Aspect ratio 0.3; RMS SpmaskErr 3.8

e) Spectr. mask area 0.275; Aspect ratio 0.45; RMS SpmaskErr 3.8

Figure 7. Test image "BloodVessels512 (a) , and image spectrum MSED-zone (white dots) along with the borders (white lines) of the rectangular, triangular and oval shapes with different shape parameters that approximate it ( b) –e).

As one can see, this MSED-zone exhibits a considerable anisotropy, which apparently evidences a certain prevalence of horizontally oriented edges of blood vessels shown in the test image. Solid lines in these images represent borders of four different standard shapes (rectangle, triangle, and two ovals) that are chosen to approximate the spectrum MSED-zone. These shapes have different aspect ratios (0.35, 0.25, 0.3, and 0.45) but all permit image reconstruction with approximately the same RMSEs (4.1, 3.7, 3.8 and 3.8) as the RMSE for the image spectrum MSED-zone. Therefore they are practically equivalent as approximations of the given spectrum MSED-zone.

The reason why no fine adjustment of shape parameters is needed for choosing spectra MSED-zones approximating shapes lies in the experimental fact that borders of image spectra MSED-zones are quite fuzzy, which one can easily see on the presented example.

The reason why areas of standard shapes always exceed areas of spectra MSED-zones they approximate is also almost obvious. Image spectra MSED-zones are composed of the largest image spectral components that reconstruct the image with a given MSE. Shapes that approximate the MSED-zones will certainly contain some quantity of "no-MSED"-zone components that by definition have lower energy than the largest ones, which form the MSED-zone, and may not contain some MSED-zone components. Therefore, in order to secure the given image reconstruction MSE, areas of MSED-zone approximating shapes must exceed areas of the corresponding MSED-zones. Therefore image sampling rate equal to the area of the MSED-zone approximating shapes, being minimal for the given approximating shape, will always to a certain degree exceed the minimal sampling rate defined by the area of the image proper MSED-zone. This sampling redundancy is the price for not knowing exact positions of spectral components that form the image spectrum MSED-zone. For the example presented in Figure 7 this redundancy is 1.67.

- Positioning image samples. As was mentioned, DCT as the sparsifying transform imposes no limitations on positions of image samples and they can be arbitrary.

- Methods of image reconstruction. There are two options for implementing image reconstruction:

• The direct matrix inversion according to Eq. 10 for computing, from available $M$ signal samples, $M$ transform coefficients chosen for the reconstruction. The found $M$ transform coefficients supplemented with the rest $N-M$ coefficients set to zero are then used for reconstruction of all required $N$ signal samples by the inverse transform. Practical usage of this option is limited because the matrix inversion is a very time consuming computational task and no fast matrix inversion algorithms are known.

• The iterative Gerchberg-Papoulis type algorithm, in which at each iteration step: (i) spectrum of the current estimate of the reconstructed image in the chosen transform is computed; (ii) all transform coefficients outside the chosen image spectrum MSED-zone approximating shape are zeroed; (iii) the modified in this way image spectrum is inversely transformed and available image samples are restored in the reconstructed image producing its estimate for the next iteration. Reconstruction iterations start from an image, in which not available samples are obtained by interpolation from the available ones using one or another interpolation method.

The above reasonings imply that, assuming DCT as the image sparsifying transform and a square sampling lattice, image sampling and reconstruction should be performed in the following steps ([14],[ 15]):

• Choose a required image spatial resolution $SpR$ (in "dots per inch)" in the same way as it is being done in the ordinary sampling.

• Given the physical dimensions $SzX$ and $SzY$ of the image (in inches) in $X$ and $Y$ image coordinates, correspondingly, determine $X/Y$ dimensions of the square sampling lattice $N_x = SpR \times SzX$ and $N_y = SpR \times SzY$.

• Choose, on the basis of evaluation of the image, one of the standard shapes for bounding image DCT spectrum MSED-zone and set its geometrical parameters.

• Inscribe the chosen shape into the rectangle of $N_x \times N_y$ samples as tightly as possible and evaluate the fraction $SS$ of the area, which the shape occupies in the rectangle (spectrum sparsity).

• Find the number of image samples $M = SS \times N_x \times N_y$ to be taken.

• Sample the image in $M$ positions distributed as uniformly over the sampling lattice of $N_x \times N_y$ samples as possible.

As one can see, the described sampling protocol is almost identical to the ordinary standard 2D sampling protocol except that in the suggested method image is sparsely sampled in a sub-set of nodes of the ordinary square sampling lattice and setting the spectrum bounding shapes for approximating the image spectra MSED-zones is required.

For image reconstruction, apply to the sampled image one of the above-mentioned image reconstruction options using, for image spectrum bounding, the chosen spectrum MSED-zone approximating shape. As a result, an image with spectrum bounded by the chosen MSED-zone approximating shape, or a bounded spectrum (BS-) image, will be obtained, which has the prescribed spatial resolution $SpR$.

Inasmuch as in the described method the sampling rate equals the area of the chosen spectrum bounding shape, the method reaches the minimal rate for the given spectrum bounding shape. However, as mentioned previously, the latter is somewhat larger than the area occupied by the actual image spectrum MSED-zone, which the chosen spectrum bounding shape approximates. Therefore, for each particular image, the method has a residual sampling redundancy equal to the ratio of the area of the chosen spectrum bounding shape to the area of the actual image spectrum MSED-zone.

In view of the said the described image sampling and reconstruction method is called the Arbitrary Sampling and Bounded Spectrum Reconstruction (ASBSR-) method.

The described ASBSR-method was extensively verified on a considerable amount of various test images ([14]). An illustrative example of reconstruction of one of the test images sampled over the uniform sampling lattice with random jitter is shown in Figure 8.

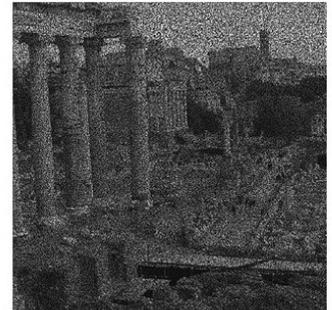

Sampled band-limited test image Sampl. redund. factor1
Sampling rate=0.505; Sampling grid: Uniform with jitter

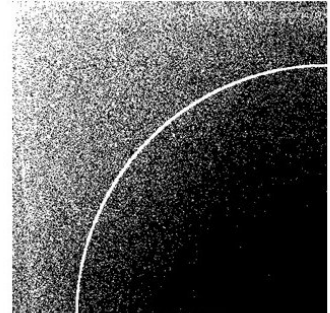

Sparse spectrum; RMS Sprs Err=5.23
Spectr Sparsity 0.314 at level JPG RMSerr 5.23.
Aspect ratio=1; EC-zone area 0.505; RMS BS Err 5.2

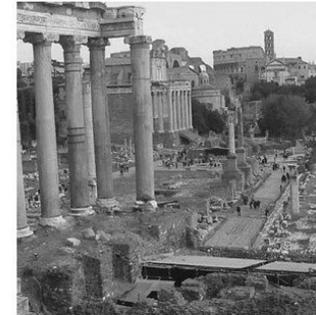

Reconstructed image; Sampling rate 0.505
RMS RecErr=1.82; RMS RecErr0=5.51; Sampl.redund 1.61

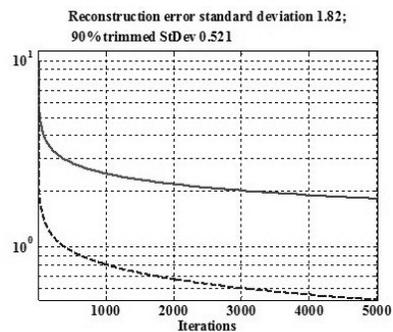

Figure 8. An illustrative example of results of experiments on image sampling and reconstruction using the ASBSR-method. From top to bottom: sampled test image "Rome512"(grey dots); image spectrum MSED-zone (white dots) and borders of its chosen approximating shape (white solid line); test image reconstructed using the iterative reconstruction algorithm; plots of root mean square of all (solid line) and of the smallest 90% (dash line) reconstruction errors vs. the number of iterations.

The experiments confirm that images sampled with sampling rates equal to the minimal rate for their chosen MSED-zone approximating shapes can be reconstructed with a sufficiently good accuracy comparable with that of image JPEG compression. The redundancy in the number of the required samples associated with the redundancy of the standard shapes approximating image spectra MSED-zones was in the experiments in the range 1.5-1.7, which is noticeably lower than the mentioned above range 2.5-5 for the sampling redundancy of compressed sensing methods.

## 8. Some practical issues and other possible applications of the ASBSR-method

In conclusion, address some practical issues of using the ASBSR-method of image sampling and reconstruction: robustness of the method to noise in sampled data, image anti-aliasing pre-filtering, recommended sample positioning, and possible applications of the method to solving under-determined inverse problems.

In distinction from the Compressed sensing methods, the ASBSR method, being a linear one, is insensitive to noise in image signals. If input image is contaminated with additive white noise, image reconstructed from the sampled data will also contain additive noise with non-zero spectrum within the shape used for bounded spectrum image reconstruction and variance equal to the variance of the input image noise times the fraction of the area of sampling base band occupied by the spectrum bounding shape.

As dictated by the sampling theory, image pre-filtering for bounding its spectrum before sampling is necessary in order to avoid distortions of reconstructed images caused by spectra aliasing due to sampling. Conventionally such pre-filtering is carried out by apertures of light sensitive cells of digital cameras and image scanners.

The ASBSR method envisages, generally, choosing spectrum MSED-zone approximating shapes individually for each particular image. Ordinary photo sensors are not capable of implementing such choice. As a solution to this problem, the usage of synthetic multiple aperture sensors can be proposed ([ 12], [ 19]). In these sensors, several individual sub-sensors are allocated for each image sample and the desired anti-aliasing filter frequency response is synthesized by an appropriate weighted summation of outputs of individual sub-sensors. The multiple aperture sensors are especially well suited for the so called single-pixel cameras, where sampling is carried out using digital micro-mirror devices, which enable the possibility of arbitrary arrangement of sampling positions. Note that single pixel cameras are recommended for implementation of Compressed sensing methods of image acquisition ([ 7])

Choosing anti-aliasing filters individually adjusted for each particular image is advisable but not very critical. As a practical alternative, the use of a universal "all purpose" shape can be considered. As such, a "pie-sector" shape can be suggested, which fits the majority of natural images quite well.

In the experimental verification of the method three types of sampling lattices were tested ([ 12], [ 13], [14], [ 15]):

- "quasi-uniform" sampling lattice, in which image samples are distributed uniformly in both image coordinates with an appropriate rounding off their positions to the nearest nodes of the dense square sampling lattice allocated for the sampled image;

- uniform sampling lattice with pseudo-random jitter, in which sample positions in both image coordinates are randomly chosen within the primary uniform sampling intervals independently in each of two image coordinates;

- the totally pseudorandom sampling lattice, in which sample positions are randomly placed with uniform distribution at nodes of the dense sampling lattice allocated for the sampled image.

For all test images used in the experiments, root mean square of reconstrgruction errors (RMSE) decayed with iterations most rapidly for the case of sampling over the uniform sampling lattices with pseudo-random jitter. Reconstruction RMSEs for totally random samplimg lattice were about 1.5-2 times and for "quasi-uniform" sampling lattice 2-2.5 times larger than those for the "uniform with jitter" sampling lattices for the same number of iterations. When "quasi-uniform" sampling lattices were used, stagnation of the iteration process was observed. This phenomenon can apparently be attributed to the emerging of regular patterns of thickening and rarefaction of sampling positions due to rounding off their coordinates to the node positions of the regular uniform sampling lattice ([ 12], [ 13],[14], [ 15]).

As already mentioned, the task of reconstruction of images of $N$ samples from $M < N$ sampled data is a special case of the under-determined inverse imaging problems. The found solution of this task, the bounded spectrum (BS-) image reconstruction, can be used for solving other under-determined inverse imaging problems as well. In Refs.[ 12], [ 13], [14], [ 15], one can find demonstrations of using this option for

- demosaicing color images;
- image super-resolution from multiple chaotically sampled video frames;
- image super-resolution in computed tomography;
- image reconstruction from their sparsely sampled Fourier spectra;
- image reconstruction from the modulus of its Fourier spectrum.